\documentclass[twocolumn,aps,prl]{revtex4-2}

\usepackage{amsfonts}
\usepackage{amssymb}
\usepackage{amsmath}
\usepackage{bm,siunitx}
\usepackage{color}
\usepackage{epsfig}
\usepackage{ifthen}
\usepackage{graphicx}
\usepackage{multirow}
\usepackage{bigstrut}
\usepackage{rotating} 
\usepackage{longtable}
\usepackage{soul}
\usepackage{xcolor}
\usepackage{comment}

\sethlcolor{yellow}

\def\be{ \begin{equation} }
\def\ee{ \end{equation} }
\def\bea{ \begin{eqnarray} }
\def\eea{ \end{eqnarray} }
\def\bse{ \begin{subequations} }
\def\ese{ \end{subequations} }

\def\P01{P_{\,\text{0}\rightarrow \text{1}} }

\def\section#1{\textbf{#1}}
\def\bt{\begin{tabular}}
\def\et{\end{tabular}}

\begin{document}

\author{Ivo S. Mihov}
\affiliation{Center for Quantum Technologies, Department of Physics, Sofia University, 5 James Bourchier blvd, 1164 Sofia, Bulgaria}
\author{Nikolay V. Vitanov}
\affiliation{Center for Quantum Technologies, Department of Physics, Sofia University, 5 James Bourchier blvd, 1164 Sofia, Bulgaria}
\title{Defying Conventional Wisdom in Spectroscopy: Power Narrowing on IBM Quantum}
\date{\today }

\begin{abstract}   
Power broadening --- the broadening of the spectral line profile of a two-state quantum transition as the amplitude of the driving field increases ---  is a well-known and thoroughly examined phenomenon in spectroscopy. 
It typically occurs in continuous-wave driving when the intensity of the radiation field increases beyond the saturation intensity of the transition.
In pulsed-field excitation, linear power broadening occurs for a pulse of rectangular temporal shape. 
Pulses with smooth shapes are known to exhibit much less power broadening, e.g. logarithmic for a Gaussian pulse shape.
It has been predicted, but never experimentally verified, that pulse shapes which vanish in time as $\sim |t|^{-\lambda}$ 
should exhibit the opposite effect --- power narrowing --- in which the post-pulse transition line width \textit{decreases} as the amplitude of the driving pulse increases.
In this work, power narrowing is demonstrated for a class of powers-of-Lorentzian pulse shapes on the IBM Quantum processor ibmq\_manila.
Reduction of the line width by a factor of over 10 is observed when increasing the pulse area from $\pi$ to $7\pi$, in a complete reversal of the power broadening paradigm.
Moreover, thorough study is conducted on the truncation of the pulse wings which introduces a (small) power-broadened term 
which prevents power narrowing from reaching extreme values. 
In the absence of other power broadening mechanisms, 
Lorentzian pulses truncated at sufficiently small values can achieve as narrow line profiles as desired.
\end{abstract}

\maketitle


Power broadening is one of the basic paradigms in spectroscopy \cite{Citron1977,Harth1995,Foot2005,Milonni2010,Roy2011,Levine2012,Bettles2020}:
when the intensity of the continuous-wave radiation field driving a two-state quantum transition increases beyond its saturation value, the excitation line width increases in proportion to the field amplitude.
This effect, which is detrimental to high-resolution spectroscopy, supplements other unwanted broadening mechanisms, such as natural broadening, Doppler broadening, collisional broadening, etc.

In pulsed-field excitation, power broadening depends on the shape of the excitation pulse and the measurement method \cite{Vitanov2009}, e.g. if the signal is collected during the excitation or after it (post-pulse). 
In particular, the post-pulse excitation line profile depends very strongly on the pulse shape.
To this end, it is well known that linear power broadening occurs for a pulse of rectangular temporal shape.
However, for smooth pulse shapes power broadening changes dramatically. 
Pulses of the ubiquitous Gaussian shape generate excitation profiles which feature weak, logarithmic power broadening~\cite{Berman1998,Vitanov2001,Halfmann2003,Vasilev2004}.
Hyperbolic-secant pulses show no broadening whatsoever: the width of the excitation line profile is the same for any pulse amplitude~\cite{Rosen1932}.

In some situations power broadening is desirable and has been used to an advantage \cite{Lindenfelser2017, Park2011, Levine2009, Calderon2008, Berry2006, Tochitsky2001}.
However, in the vast majority of applications, most notably in high-resolution spectroscopy, it is detrimental and various ideas for suppressing it have been explored over the years. 
Examples include replacing continuous by pulse driving and post-pulse measurement of excitation 
\cite{Conde2006,deGroote2013,Sonnenschein2014,deGroote2015,deGroote2017,Koszorus2019,Kron2020, Dreau2011, Dreau2012, Jamonneau2016, Zhang2021, Yang2022, Zhang2022}, and
sophisticated setups involving multiple states and paths and using quantum interference
\cite{Dudovich2005, Chalupczak2010, Zanon-Willette2011, Florez2013, Niemeyer2013, Waters2015, Offer2016, Reimer2016, Li2017, Jamali2021, Prajapati2021, Zhu2021, Gawlik2022, Lukin1997, Erhard2001, Deb2009, Xiao2009, Finkelstein2019, Xiao2023}.

In a complete reversal of the power broadening paradigm, it has been predicted, but never verified in an experiment, that for a family of pulse shapes with wings vanishing polynomially in time, the spectral line must exhibit the opposite effect --- power \textit{narrowing}~\cite{Boradjiev2013}. 
In particular, for the Lorentzian pulse shape, the line width must scale as $1/\Omega_0$, the inverse of the peak Rabi frequency.
Moreover, it has been predicted that power narrowing can be observed for pulse shapes of any power of Lorentzian higher than $\frac12$: the closer the power to $\frac12$ the stronger the power narrowing effect. 

In this work, we demonstrate power narrowing, by as much as an order of magnitude, for pulses shaped like powers of Lorentzian, obtained with  ibmq\_manila, one of IBM's open quantum processors. 
The effect is only limited by the inevitable truncation of the pulse wings, an impediment which is studied in detail.


Remarkably, the degree of power broadening or narrowing of the post-pulse excitation spectral profile can be described in a simple fashion by using the concept of adiabatic evolution \cite{Born1928,Boradjiev2013}, see Fig.~\ref{fig:mechanism}(left).
The method is summarized in the Supplemental Material~\cite{SM}.
It has been predicted \cite{Boradjiev2013} that for a class of pulses whose tail falls off as $f(t)\sim 1 / |t|^{\lambda}$ with $\lambda>1$, $|t|\rightarrow \infty$, we must have 
\be
    \Delta_{\frac12} \tau \propto \left(\Omega_0 \tau\right)^{-\frac{1}{\lambda-1}},
    \label{eq-deltamlorentzian}
\ee
where $\tau$ is the pulse width and $\Delta_{\frac12}$ is the spectral line width at half maximum.
Here the detuning $\Delta = \omega_0-\omega$ is the difference between the qubit transition frequency $\omega_0$ and the carrier frequency of the driving pulsed field $\omega$, while the Rabi frequency $\Omega(t)=\Omega_0 f(t)$ measures the magnitude of the field-qubit interaction.
Equation \eqref{eq-deltamlorentzian} shows that as the pulse amplitude $\Omega_0$ increases, $\Delta_{\frac12}$ \textit{decreases}, or in other words --- we have \textit{power narrowing}. 
A family of pulse shapes that satisfy this requirement are the powers of the Lorentzian function ($\lambda=2n$),
\be
    L^n(t) = \frac{1}{\left[1+(t/\tau)^2\right]^n}\quad \left(n>\tfrac12\right),
    \label{eq-lorentzian}
\ee
which we use here with powers from $n=\frac35$ to $2$.

\begin{figure}[tb]
\bt{cc}
\includegraphics[width=0.5\columnwidth]{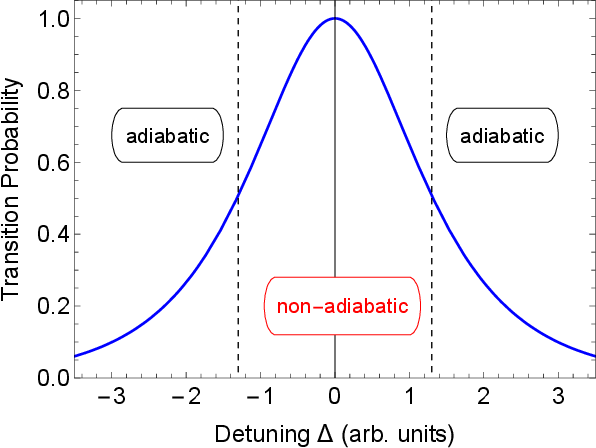}  
\includegraphics[width=0.5\columnwidth]{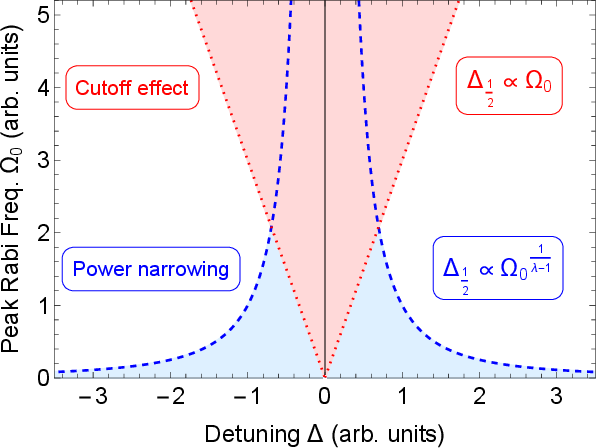}  
\et
\caption{
    \textit{Left:} A typical spectral profile of the post-pulse transition probability between two quantum states. The width of the spectral line profile is determined by the detuning range wherein the adiabatic condition is violated.
    \textit{Right:} Qualitative behavior of the transition probability excitation landscape vs detuning and peak Rabi frequency (on a color map). The dashed lines depict the limits on the spectral line width induced by the power narrowing effect due to the Lorentzian pulse shape. The area between the dotted lines show the power broadened contribution due to the truncation of the pulse wings. 
  }
  \label{fig:mechanism}
\end{figure}

In this work we use qubit 0 of ibmq\_manila, one of the IBM Quantum Falcon r5.11L Processors~\cite{IBM,Koch2007}. 
With the $L^{\frac{3}{2}}$ pulse, the T1 and T2 coherence times were $148.16~\mathrm{\mu s}$ and $56.53~\mathrm{\mu s}$, while with the other pulses they were $166.44~\mathrm{\mu s}$ and $116.53~\mathrm{\mu s}$.
The rest of the parameters are described in the Supplemental Material~\cite{SM}.
The T2-limited linewidth is $1\,/\,T2 \approx 10-20~$kHz, compared to linewidths of $5-100~$MHz, suggesting that dephasing effects are negligible.

We used the Qiskit Pulse framework for Python to establish pulse-level control of the hardware. 
The most relevant limitations of the Qiskit Pulse framework are:
(i) a (soft) upper limit on the pulse duration at few $\mu s$;
(ii) a lower limit on the pulse duration due to discretisation of the pulse amplitude shape into small intervals of $2/9$~ns each;
(iii) an upper limit of the magnitude of the driving pulse;
(iv) default measurement of the population of the first two levels only, where leakage is included in the excited level.
Each point on the figures below is the average of 1024 shots.

In order to maximize power narrowing, the power $n$ of the shape $L^n(t)$ of Eq.~\eqref{eq-lorentzian} must be as close to $\frac12$ as possible. 
However, as $n\to\frac12$ the pulse becomes very long and we need to cut the pulse wings off at certain times $\pm t_c$ (chosen symmetric). 
The truncation gives rise to two sudden jumps in the pulse shape which generate $\delta$-function features in the nonadiabatic coupling $\dot\vartheta(t)$, inducing an additional term in the transition probability \cite{Boradjiev2013pra}. 
A simple calculation gives the amplitude of the cut-off-induced term  (neglecting oscillations) as
\be\label{P-truncation}
P_c \approx \frac{\Omega_c^2}{\Omega_c^2+\Delta^2} [1-\P01(\Delta)] ,
\ee
where $\Omega_c=\Omega(t_c)$ and $\P01$ is the ideal transition probability (with no truncation).
This cut-off artefact $P_c$
exhibits linear power broadening $\Delta_{\frac12} \sim \Omega_c $, as depicted in Fig.~\ref{fig:mechanism}(right), and hence it hinders the desired power narrowing effect.
Its effect will be pronounced if the truncation is severe and $\Omega_c$ is large, but it will have lesser effect when the truncation is far enough and $\Omega_c$ is small.

The truncation effect tends to increase for small $\Delta$; however, near resonance its effect is mitigated by the factor $1-\P01(\Delta)$, which should be very small if $\P01(0)\approx 1$.
Formula \eqref{P-truncation}, applied at $\Delta_{\frac12}$, shows that the cut-off artefact affects the spectral line profile when $\Omega_c\gtrsim\Delta_{\frac12}$, and suggests how large an $\Omega_c$ we can afford. 

\begin{figure}[tb]
\bt{c}
  \includegraphics[width=1\columnwidth]{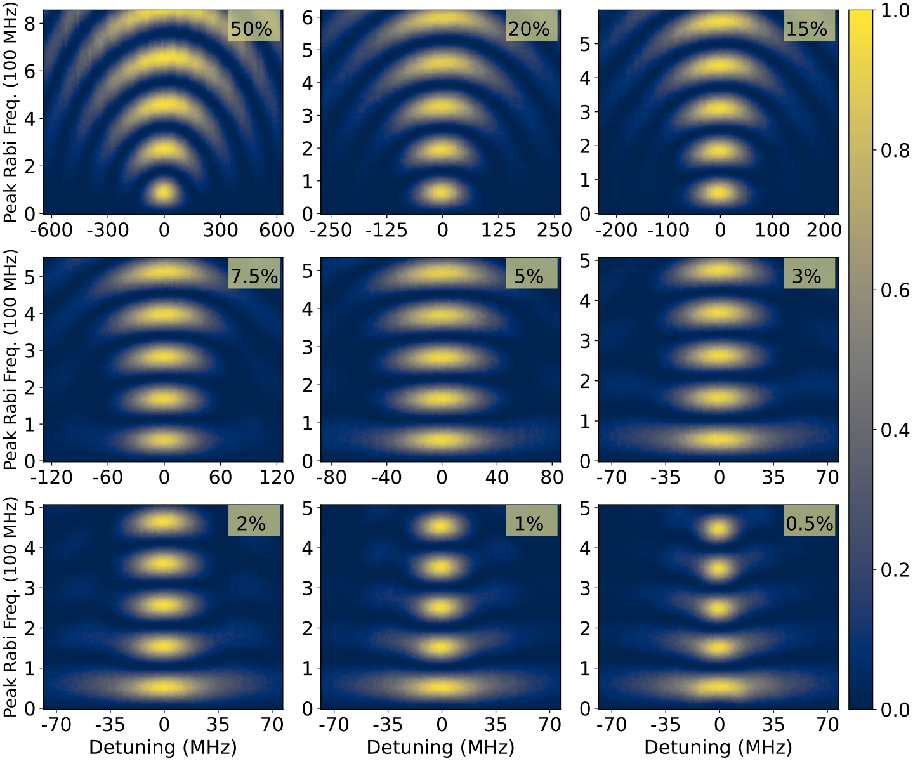}  
\et
  \caption{Excitation landscapes (transition probability vs detuning $\Delta$ and peak Rabi frequency $\Omega_0$) of a Lorentzian pulse truncated at the cut-off points $\pm t_c$, at which the pulse amplitude value $\Omega_c = \Omega_0 f(t_c)$ is a fraction of its maximum value $\Omega_0$, indicated in each frame.
  }
  \label{figs-from-brd-to-nrw}
\end{figure}

In order to explore the truncation effect we use the Lorentzian pulse shape ($n=1$) with width $\tau=21.33$~ns.
Figure~\ref{figs-from-brd-to-nrw} presents the map of population transfer between the two states in a set of 9 figures obtained 
by moving the truncation points from the pulse center to its wings, from cut-off Rabi frequency values $\Omega_c=50\%\Omega_0$ to $0.5\%\Omega_0$.
Along the vertical line of exact resonance ($\Delta=0$) in all plots we see the usual Rabi oscillations, with peaks at pulse areas of $\pi$, $3\pi,\ldots,9\pi$. 
Off resonance, however, the excitation landscape changes very strongly from plot to plot. 
When the truncation points are close to the center, the excitation map looks very similar to the one of a rectangular pulse, with typical power broadening, see Fig.~\ref{figs-from-brd-to-nrw} (top row).
The power narrowing effect (the decrease of the widths of the bright domains as the Rabi frequency amplitude increases) appears when we truncate the pulse at less than 2\% of its amplitude, see Fig.~\ref{figs-from-brd-to-nrw} (bottom row).
A curious limiting case is the truncation at 3\%: power narrowing is observed from the first to the third peak, which is then overtaken by power broadening seen in the width increase from the third to the fifth peak.
Indeed, similar behavior should be seen in the other cases of small truncation levels (bottom row), if we were able to go to higher Rabi frequency.

We notice a few other trends in Fig.~\ref{figs-from-brd-to-nrw}. 
Most importantly, the lowest bright spot shows no significant line width change, situated between $\pm 35$~MHz.
However, all other peaks exhibit dramatic narrowing as the duration increases and truncation gets smaller.
We also note the transformation of the high-excitation stripes from concave to convex shape as we move from power broadening to power narrowing.

\begin{figure}[tb]
   \centering
\bt{c}    
    \includegraphics[width=1\columnwidth]{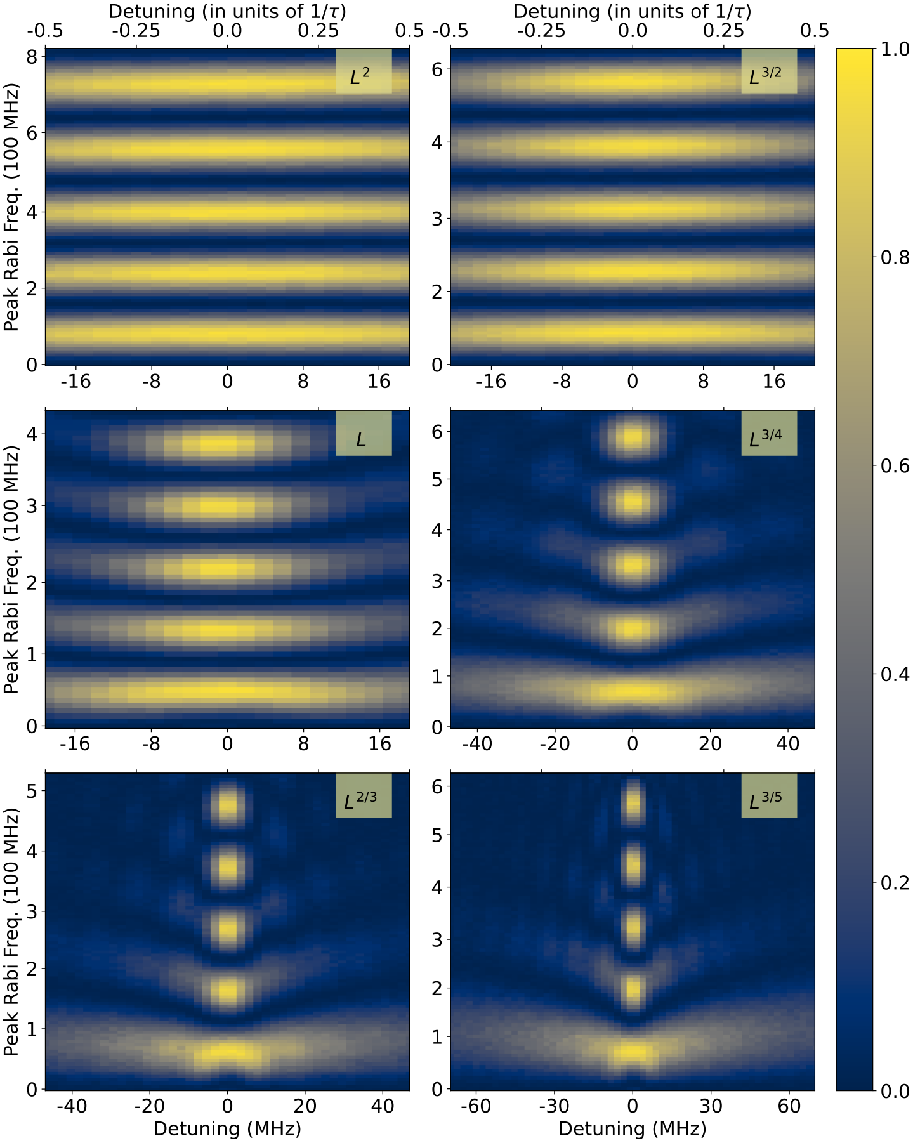}
\et
    \caption{
    Excitation landscapes (transition probability vs detuning $\Delta$ and peak Rabi frequency $\Omega_0$) of different pulse shapes $L^n(t)$, with the power $n$ listed in each plot.
       }
    \label{fig-narrowing-2d}
\end{figure}

%
We now turn our attention to pushing the power narrowing effect as far as possible.
To this end, we used 6 different pulse shapes $L^n(t)$ described by Eq.~\eqref{eq-lorentzian} with $n=2,\frac32,1,\frac34,\frac23,\frac35$, for which theory predicts the scaling behavior $\Delta_{\frac12} \propto \Omega^{-\nu}$, with $\nu = 1/(2n-1) = \frac13, \frac12, 1,2,3,5$ for the 6 selected values of $n$, respectively.
We truncated all pulse shapes at 0.5\% of their maximum value.
The excitation landscape plots are shown in Fig.~\ref{fig-narrowing-2d}.
The $L^2(t)$, $L^{\frac32}(t)$ and $L^1(t)$ pulses have the same pulse width $T=24.89$~ns.
The $L^{\frac34}(t)$ and $L^{\frac23}(t)$ pulses have a smaller pulse width, $10.67$~ns,
in order to facilitate truncation. 
The $L^{\frac35}(t)$ pulse has the smallest pulse width, $7.11$~ns, 
in order to further facilitate truncation, which becomes challenging as the power $n$ of $L^n(t)$ approaches $\frac12$. 
In order to compare the six plots with each other, we plotted all frames in the same range of the dimensionless parameter $\Delta \tau$, given in the upper horizontal axis.

The six plots in Fig.~\ref{fig-narrowing-2d} show that the effect of the power $n$ of the Lorentzian shapes $L^n(t)$ is dramatic.
The narrowing for $L^2(t)$ and $L^{\frac32}(t)$ (top row) is barely noticeable, although it exists (see Fig.~\ref{fig-narrowing-detun} below).
The other 4 plots reveal very strong power narrowing: even for $L^1(t)$ the narrowing is significant, but for $L^{\frac34}(t)$, $L^{\frac23}(t)$ and  $L^{\frac35}(t)$ the narrowing is striking even when comparing the first two maxima widths. 

%

\begin{figure}[tb]
    \centering
    \includegraphics[width=1\columnwidth]{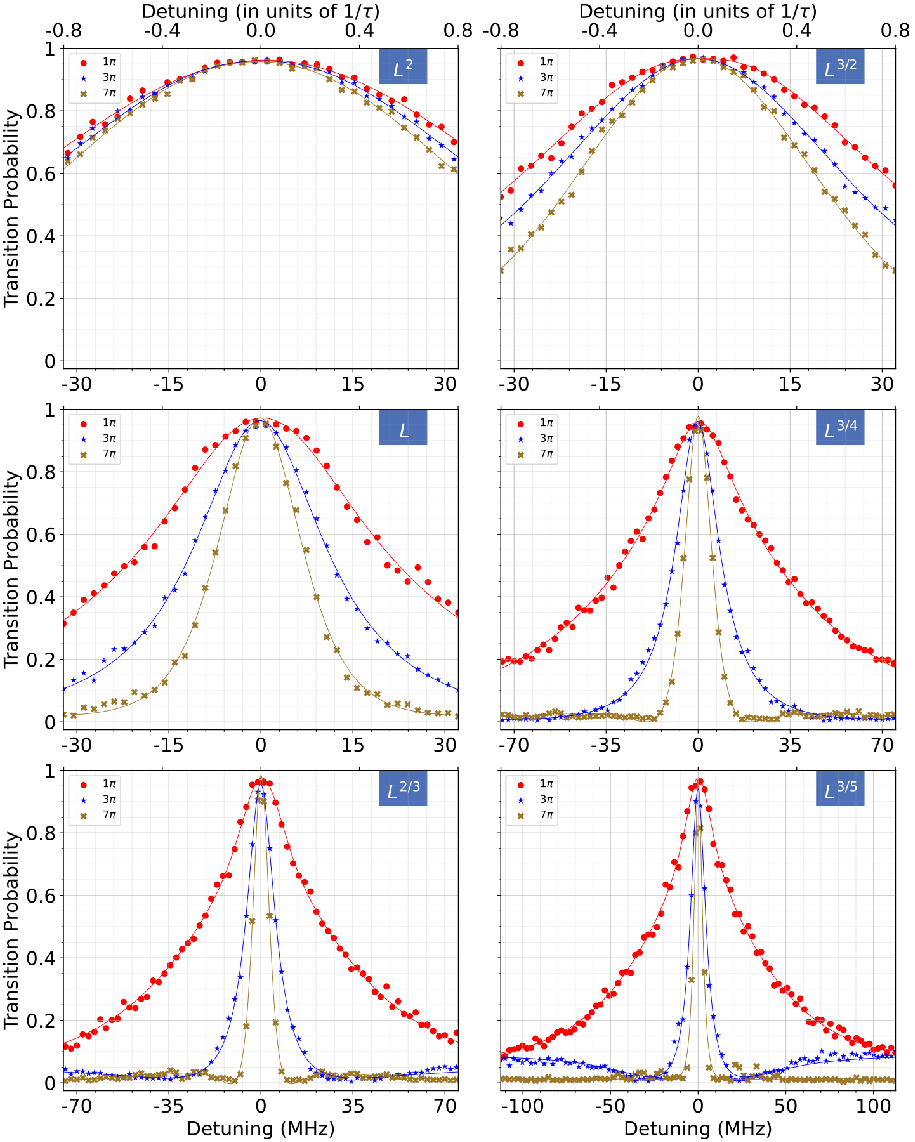}
    \caption{The transition probability vs detuning for 6 different pulse shapes $L^n(t)$. The data points are fitted with a function that is a sum of a hyperbolic secant and a Lorentzian.
     }
    \label{fig-narrowing-detun}
\end{figure}

\begin{table}[tb]
\begin{tabular}{|c|ccc|c|}
\hline
\multirow{2}{*}{$n$} & \multicolumn{3}{c|}{FWHM (MHz)} & \multirow{2}{*}{$\frac{\text{FWHM}_{\pi}}{\text{FWHM}_{7\pi}}$}  \\
 & $\pi$ & $3\pi$ & $7\pi$ &  \\ 
 \hline
2 & 98.5 & 91.7 & 82.4 & 1.20 \\
$3/2$ & 74.7 & 60.4 & 48.1 & 1.55 \\
1 & 47.2 & 26.1 & 16.2 & 2.90 \\
$3/4$ & 67.0 & 20.4 & 11.5 & 5.81 \\
$2/3$ & 52.3 & 13.0 & 6.9 & 7.57 \\
$3/5$ & 56.3 & 10.5 & 5.4 & 10.35 \\
\hline
\end{tabular}
\caption{FWHM of the profiles generated by the $L^n(t)$ pulse shapes for pulse areas $\pi$, $3\pi$ and $7\pi$, and the ratio of the FWHM for the first and third of these. }
\label{Table:FWHM}
\end{table}

Next, we take cuts along the horizontal and vertical axes of Fig.~\ref{fig-narrowing-2d} in order to examine the power narrowing effect in a quantitative manner.
Figure~\ref{fig-narrowing-detun} shows horizontal cuts across three exact maxima in Fig.~\ref{fig-narrowing-2d} at pulse areas of $\pi$, $3\pi$ and $7\pi$.
The other pulse specifications are the same as in Fig.~\ref{fig-narrowing-2d}. 
Table \ref{Table:FWHM} lists the measured full-width-at-half-maximum~(FWHM) for each pulse at pulse areas of $\pi$, $3\pi$ and $7\pi$.
Power narrowing is observed for all pulse shapes and all pulse areas.
For $n=2$, the difference between the $\pi$, $3\pi$ and $7\pi$ curves is marginal, albeit in the direction of narrowing, with some 20\% FWHM narrowing from $\pi$ to $7\pi$. 
This difference increases for $n=\frac32$, with some 55\% narrowing. 
The other 4 plots show dramatic power narrowing, which gets stronger toward smaller $n$,
 with FWHM narrowing from $\pi$ to $7\pi$ by factors of 2.90, 5.81, 7.57 and 10.35 for $n=1,\frac34,\frac23,\frac35$, respectively.

\begin{figure}[tb]
    \centering
    \includegraphics[width=1.0\columnwidth]{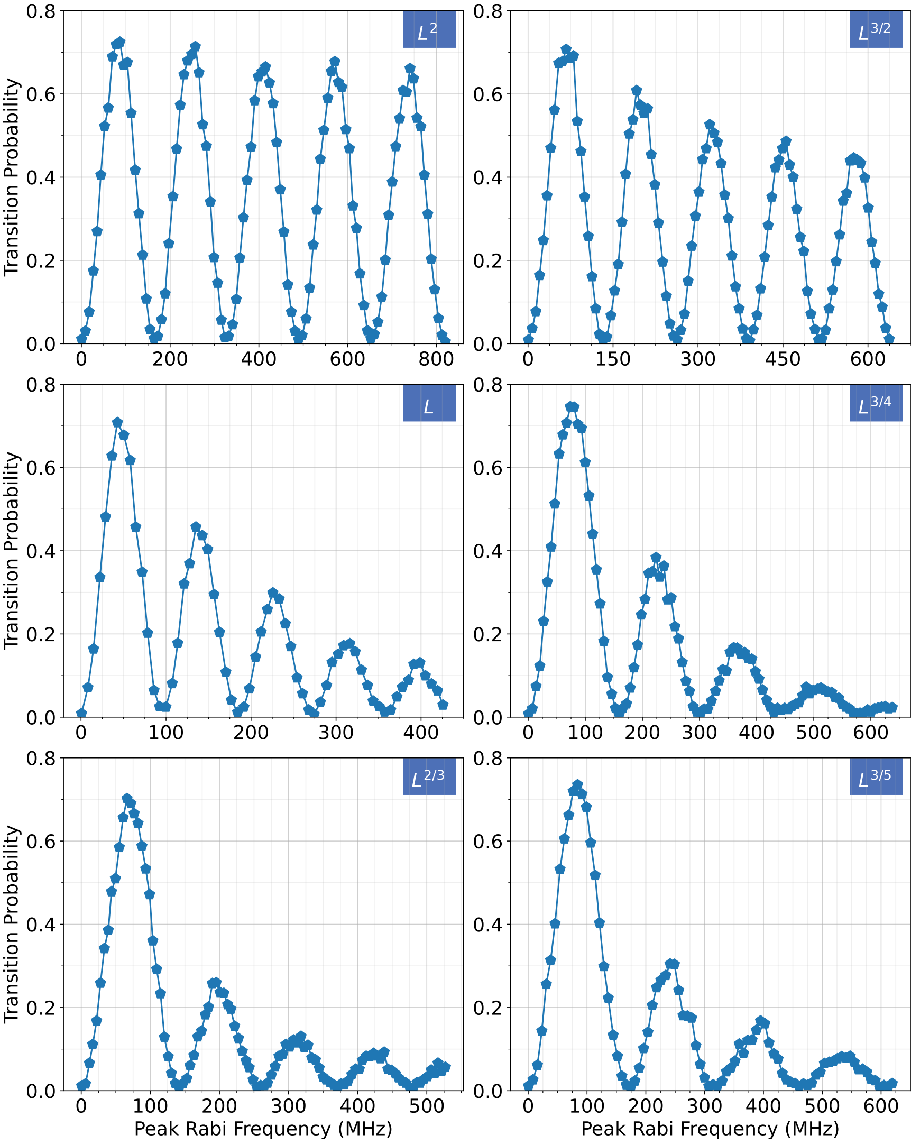}
    \caption{Off-resonant Rabi oscillations of 6 Lorentzian-based pulses. The 6 plots are vertical slices of the colour maps in Fig.~\ref{fig-narrowing-2d} at $\Delta = 30$, $25$, $12.5$, $12.5$, $12.5$ and $12.5$~MHz for the $n=2$, $3/2$, $1$, $3/4$, $2/3$ and $3/5$ Lorentzian pulses respectively, labelled in the upper-right corner of each plot.}
    \label{fig-narrowing-rabi-osc}
\end{figure}

The power narrowing effect, although very pronounced, is less than what the simple adiabatic theory predicts. 
This discrepancy is rooted in the cut-off broadening effect, which sets a lower limit of the order of $\Omega_c$ on the line width.
Had we been able to cut-off at less than 0.5\%, say at $\epsilon$\%, we would see further power narrowing improvement by a factor of $(0.5/\epsilon)^2$.
Additionally, leakage could be hindering our ability to observe the power narrowing extent in $L^{2}$ and $L^{\frac{3}{2}}$ pulses. Our simulations estimate that leakage is significant ($\approx 10\%$) in $L^{2}$ and $L^{\frac{3}{2}}$ pulses, and negligible ($<1\%$) in the other 4 pulses.

Furthermore, the spectral resolution of the $L^{\frac{3}{5}}(t)$ pulse was numerically compared to traditional Ramsey measurement and found to be almost on par. Further research would be needed for a conclusive result.

Figure~\ref{fig-narrowing-rabi-osc} shows vertical slices of the transition landscape maps of Fig.~\ref{fig-narrowing-2d} at certain fixed detuning. 
In all plots, we observe off-resonant Rabi oscillations with pronounced amplitude damping, as expected from Fig.~\ref{fig-narrowing-2d}.
For $L^2(t)$ the damping is weakest but still present. 
For $L^1(t)$ the damping is already very strong and it gets even stronger toward $L^{\frac35}(t)$.
This power damping can be considered as an indirect evidence of power narrowing, as evident from Fig.~\ref{fig-narrowing-2d}.
Such damping has been observed earlier \cite{Conover2011} but has not been linked to power narrowing.

In conclusion, we demonstrated, on one of IBM Quantum processors, that the spectral line profiles of a qubit excited by microwave pulses of Lorentzian temporal shape and its powers experience power narrowing.
This phenomenon defies the conventional wisdom in spectroscopy which stipulates power broadening of spectral lines.
The physical mechanism of this unusual phenomenon is rooted in the basic concept of adiabatic population return. 

The observations on ibmq\_manila quantum processor exhibit pronounced power narrowing for pulses of temporal shape $L^n(t)$ ($n>\frac12$), with the effect getting stronger as $n$ decreases, reaching an order of magnitude for $n=\frac35$.
The extent of power narrowing was limited by the inevitable truncation of the wings, for which an explicit analytic formula has been derived.
Following an extensive study we concluded that the wings should be cut off at less than 1\% of the maximum central value of the Rabi frequency for the effect to be observed, otherwise it is overwhelmed by the cut-off-induced power broadening. 
In experiments which allow for cut-off at 0.1\% or less, power narrowing by a factor of 100 to 1000 should be feasible. 

These results open exciting avenues for significant improvements in high-resolution spectroscopy, with far-reaching implications along numerous directions of research.


\acknowledgments 

This research is supported by the Bulgarian national plan for recovery and resilience, contract BG-RRP-2.004-0008-C01 (SUMMIT), project number 3.1.4  and by the European Union’s Horizon Europe research and innovation program under Grant Agreement No. 101046968 (BRISQ). We acknowledge the use of IBM Quantum services for this work. The views expressed are those of the authors, and do not reflect the official policy or position of IBM or the IBM Quantum team. 


\end{document}